\documentclass[11pt]{article}
\usepackage{amsmath,amssymb}
\newcommand{\bea}{\begin{eqnarray}}
\newcommand{\eea}{\end{eqnarray}}

\newcommand{\de}{\mathrm{d}}

\newcommand{\im}{\mathrm{i}}

\newcommand{\ft}[2]{{\textstyle\frac{#1}{#2}}}

\newsavebox{\uuunit}
\sbox{\uuunit}
    {\setlength{\unitlength}{0.825em}
     \begin{picture}(0.6,0.7)
        \thinlines
        \put(0,0){\line(1,0){0.5}}
        \put(0.15,0){\line(0,1){0.7}}
        \put(0.35,0){\line(0,1){0.8}}
       \multiput(0.3,0.8)(-0.04,-0.02){10}{\rule{0.5pt}{0.5pt}}
     \end {picture}}
\newcommand {\unity}{\mathord{\!\usebox{\uuunit}}}

\numberwithin{equation}{section}

\begin{document}
\begin{titlepage}
\begin{center}
\hfill LMU-ASC 84/06 \\
\hfill ITP-UU-06/54 \\
\hfill SPIN-06/44   \\
\hfill {\tt hep-th/0612225}\\
\vskip 6mm

{\Large \textbf{Black hole entropy functions and attractor equations }}
\vskip 8mm

\textbf{G. L.~Cardoso$^{a}$, B. de Wit$^b$ and S.~Mahapatra$^{c}$}

\vskip 4mm
$^a${\em Arnold Sommerfeld Center for Theoretical Physics\\
Department f\"ur Physik,
Ludwig-Maximilians-Universit\"at M\"unchen, Munich, Germany}\\
{\tt gcardoso@theorie.physik.uni-muenchen.de}\\[1mm]
$^b${\em Institute for Theoretical Physics} and {\em Spinoza
  Institute,\\ Utrecht University, Utrecht, The Netherlands}\\
{\tt  B.deWit@phys.uu.nl} \\[1mm]
$^c${\em 
Physics Department, Utkal University, 
Bhubaneswar 751 004, India}\\
{\tt swapna@iopb.res.in}\\[1mm]

\end{center}
\vskip .2in
\begin{center} {\bf ABSTRACT } \end{center}
\begin{quotation}\noindent
  The entropy and the attractor equations for static extremal black
  hole solutions follow from a variational principle based on an
  entropy function. In the general case such an entropy function can
  be derived from the reduced action evaluated in a near-horizon
  geometry. BPS black holes constitute special solutions of this
  variational principle, but they can also be derived directly from a
  different entropy function based on supersymmetry enhancement at the
  horizon. Both functions are consistent with electric/magnetic
  duality and for BPS black holes their corresponding OSV-type
  integrals give identical results at the semi-classical level. We
  clarify the relation between the two entropy functions and the
  corresponding attractor equations for $N=2$ supergravity theories
  with higher-derivative couplings in four space-time dimensions. We
  discuss how non-holomorphic corrections will modify these entropy
  functions.
\end{quotation}

\vfill
\end{titlepage}
\eject
\section{Introduction}
\setcounter{equation}{0}
An important feature of (static) extremal black hole solutions is that
scalar fields (often called moduli) tend to fixed values at the
horizon determined by the black hole charges. These values are
independent of the asymptotic values of the fields at spatial
infinity. This fixed point behaviour is encoded in so-called attractor
equations, which, in the generic case, can be understood from the
field equations associated with the reduced action taken at a Killing
horizon. The attractor equations are a crucial ingredient in comparing
the macroscopic (or field-theoretic) black hole entropy with the
microscopic (or statistical) entropy of a corresponding brane
configuration. This and corresponding aspects of the relation between
classical and quantum black holes have been studied extensively in the
context of $N=2$ supergravity in four space-time dimensions.
Especially for BPS black holes many important results have been
obtained. The inclusion of higher-derivative interactions into the
effective actions often played a crucial role. For BPS black holes the
attractor equations can be understood entirely from supersymmetry
enhancement at the horizon.  Obviously they must correspond to special
solutions of the more general attractor equations based on a reduced
action.

In this paper we study the relation between the more general attractor
equations and the BPS attractor equations for static extremal black
holes in four space-time dimensions. This can be done conveniently in
terms of corresponding entropy functions that form the basis of an
underlying variational principle. In the presence of higher-derivative
actions it is very difficult to explicitly construct black hole
solutions.  However, by concentrating on the near-horizon region one
can usually determine the fixed-point values directly without
considering the interpolation between the horizon and spatial
infinity. This approach was first applied to BPS black holes without
higher-derivative interactions in
\cite{Ferrara:1995ih,Strominger:1996kf,Ferrara:1996dd,Ferrara:1996um,
  Ferrara:1997tw,Gibbons:1997cc,Behrndt:1996jn,Denef:2000nb} and then
with higher-derivative interactions in
\cite{LopesCardoso:1998wt,LopesCardoso:1999ur,LopesCardoso:2000qm,
LopesCardoso:2006bg}.  It was also applied to non-BPS extremal black
holes in
\cite{Ferrara:1997tw,Gibbons:1997cc,Sen:2005wa,Goldstein:2005hq,Sen:2005iz,
  Kallosh:2005ax, Tripathy:2005qp, Giryavets:2005nf,Prester:2005qs,
  Alishahiha,Kallosh:2006bt,Chandrasekhar:2006kx,
Bellucci:2006ew,Kallosh:2006bx,Sahoo:2006rp,Alishahiha:2006jd,
Exirifard:2006qv,Sahoo:2006pm,Astefanesei:2006sy,Dabholkar:2006tb}.  In the
presence of higher-derivative interactions full interpolating
solutions have been studied for BPS black holes in
\cite{LopesCardoso:2000qm,Sen:2004dp,Hubeny:2004ji,Bak:2005mt}.

For $N=2$ BPS black holes with higher-derivative interactions the
attractor equations follow from classifying possible solutions with
full supersymmetry \cite{LopesCardoso:2000qm}.  As it turns out
supersymmetry determines the near-horizon geometry (and thus the
horizon area), the values of the moduli fields in terms of the charges
and the value of the entropy as defined by the Noether charge
definition of Wald \cite{Wald:1993nt}.  For more general extremal black
holes the analysis is more subtle and makes use of an action principle
\cite{Sen:2005wa}.  When dealing with spherically symmetric solutions,
one can integrate out the spherical degrees of freedom and obtain a
reduced action for a $1+1$ dimensional field theory. This action still
describes the full black hole solutions. Under certain conditions the
fixed values at the horizon can be obtained by considering the reduced
action in a $1+1$ dimensional near-horizon geometry which has an
enhanced symmetry (usually one has $AdS_2$).  Near the horizon other
fields respect this symmetry as well (when the enhanced symmetry is
maximal the fields are all covariantly constant), so that the
two-dimensional integral in the reduced action can be dropped and one
obtains a potential depending on variables that specify the values of
the fields at the Killing horizon. Actually the number of relevant
variables can often be reduced already at an earlier stage by imposing
some of the equations of motion at the level of the interpolating
solution, but this represents no problem of principle.

This paper is organized as follows. In
section~\ref{sec:entropy-functions} we consider the entropy function,
both in the reduced action approach of \cite{Sen:2005wa} and in the
context of BPS black holes (the latter for the case of $N=2$
supergravity based on \cite{Behrndt:1996jn,LopesCardoso:2006bg}). We
discuss those features that are relevant for electric/magnetic
duality.  In
section~\ref{sec:N=2sugra} we evaluate the entropy function based on
the action of a general $N=2$ supergravity theory following
\cite{Sahoo:2006rp}, and we relate it to the BPS entropy function.  We
display the associated variational equations with and without
higher-curvature interactions. For BPS black holes both entropy
functions can be used in the definition of a corresponding duality
invariant OSV-type integral and lead to identical results at the semi-
classical level.
In section \ref{sec:discussion} we briefly comment
on corrections to the entropy functions due to other higher-derivative
interactions associated with matter multiplets.  We also discuss the
modification of the entropy functions by non-holomorphic corrections.

\section{Entropy functions}
\label{sec:entropy-functions}
\setcounter{equation}{0}
In this section we will briefly consider the entropy function derived
from the action evaluated in a near-horizon geometry for some rather
general theory and the entropy function that pertains to static BPS
black holes in $N=2$ supergravity in four space-time dimensions.  
\subsection{The reduced action and the entropy function}
\label{sec:reduced-action}
When considering spherically symmetric solutions one may integrate out
the spherical degrees of freedom. This leads to a reduced action,
which we consider here for a general system of abelian vector gauge
fields, scalar and matter fields coupled to gravity. The geometry is
then restricted to the product of the sphere $S^2$ and a $1+1$ dimensional
space-time, and the dependence of the fields on the $S^2$ coordinates
$\theta$ and $\varphi$ is fixed by symmetry arguments. For the moment
we will not make any assumption regarding the dependence on the
remaining two cooordinates $r$ and $t$.  Consequently we write the
general field configuration consistent with the various isometries as
\begin{eqnarray}
  \label{eq:general-fields}
  &&
  \mathrm{d}s^2{}_{(4)} = g_{\mu\nu} \mathrm{d}x^\mu\mathrm{d}x^\nu =
  \mathrm{d}s^2{}_{(2)} 
  + v_2 \Big(\de \theta^2 +\sin^2\theta \,\de\varphi^2\Big)\,,
  \nonumber\\ 
  &&
  F_{rt}{}^I = e^I\,,\qquad F_{\theta\varphi}{}^I = \frac{p^I}{4\pi}\,
  \sin\theta \,.
\end{eqnarray}
Here the $F_{\mu\nu}{}^I$ denote the field strengths associated with a
number of abelian gauge fields. The $\theta$-dependence of
$F_{\theta\varphi}{}^I$ is fixed by rotational invariance and the
$p^I$ denote the magnetic charges. The latter are constant by virtue
of the Bianchi identity, but all other fields are still functions of
$r$ and $t$. As we shall see in a moment the fields $e^I$ are dual to
the electric charges. The radius of $S^2$ is defined by the field
$v_2$. The line element of the $1+1$ dimensional space-time will be
expressed in terms of the two-dimensional metric $\bar g_{ij}$, whose
determinant will be related to a field $v_1$ according to, 
\begin{equation}
  \label{eq:v-1}
  v_1 = \sqrt {\vert\bar g\vert} \,.
\end{equation}
Eventually $\bar g_{ij}$ will be taken proportional to an $AdS_2$ metric,
\begin{equation}
  \label{eq:bar-g}
  \mathrm{d}s^2{}_{(2)} = \bar g_{ij} \,\mathrm{d}x^i\mathrm{d}x^j =  
  v_1\Big(-r^2\,\mathrm{d}t^2 + \frac{\de r^2}{r^2}\Big) \,.
\end{equation} 
In addition to the fields $e^I$, $v_1$ and $v_2$ there may be a number
of other fields which for the moment we denote collectively by
$u_\alpha$.

As is well known theories based on abelian vector fields are subject
to electric/magnetic duality, because their equations of motion
expressed in terms of the dual field
strengths,\footnote{
  Here and henceforth we assume that the Lagrangian depends on the
  abelian field strengths but not on their space-time derivatives.
  This restriction is not an essential one. In case that the
  Lagrangian contains derivatives of field strengths, one replaces the
  derivative of the Lagrangian in \eqref{eq:dual-F} by the
  corresponding functional derivative of the action. We also assume
  that the gauge fields appear exclusively through their field
  strengths. } 
\begin{equation}
  \label{eq:dual-F}
  G_{\mu\nu I} =  \sqrt{\vert g\vert}\,
  \varepsilon_{\mu\nu\rho\sigma}\, 
  \frac{\partial \mathcal{L}}{\partial F_{\rho\sigma}{}^I} \,,
\end{equation}
take the same form as the Bianchi identities for the field strengths
$F_{\mu\nu}{}^I$. Adopting the conventions where
$x^\mu=(t,r,\theta,\varphi)$ and $\varepsilon_{tr\theta\varphi}= 1$,
and the signature of the space-time metric equals $(-,+,+,+)$ as is
obvious from \eqref{eq:bar-g}, it follows that, in the background
\eqref{eq:general-fields},
\begin{eqnarray}
  \label{eq:G}
  G_{\theta\varphi\,I}\!&=&\!- v_1 v_2 \,\sin\theta\,
  \frac{\partial\mathcal{L}} {\partial F_{rt}{}^I}= - v_1 v_2
  \,\sin\theta\,   \frac{\partial\mathcal{L}} {\partial e^I}  \,,
  \nonumber\\ 
  G_{rt\,I}\!&=& \!- v_1v_2 \,\sin\theta\, \frac{\partial\mathcal{L}}
  {\partial F_{\theta\varphi}{}^I} = - 4\pi\,v_1v_2\,
  \frac{\partial\mathcal{L}} {\partial p^I}  \,.
\end{eqnarray}
These two tensors can be written as $q_I\,\sin\theta/(4\pi)$ and
$f_I$. The quantities $q_I$ and $f_I$ are conjugate to $p^I$ and $e^I$,
respectively, and can be written as
\begin{eqnarray}
  \label{eq:def-q-f}
  q_I(e,p,v,u) &=&- 4\pi\, v_1v_2 \,
  \frac{\partial\mathcal{L}} {\partial e^I}  \,, \nonumber \\
  f_I(e,p,v,u) &=&-  4\pi\,v_1v_2 \,
  \frac{\partial\mathcal{L}} 
  {\partial p^I}  \,. 
\end{eqnarray}
They depend on the constants $p^I$ and on the fields $e^I$, $v_{1,2}$
and $u_\alpha$, and possibly their $t$ and $r$ derivatives, but no
longer on the $S^2$ coordinates $\theta$ and $\varphi$. Upon imposing
the field equations it follows that the $q_I$ are constant and
correspond to the electric charges. Obviously our aim will be to
obtain a description in terms of the charges $p^I$ and $q_I$, rather
than in terms of the $p^I$ and $e^I$.

Electric/magnetic duality transformations are induced by rotating the
tensors $F_{\mu\nu}{}^I$ and $G_{\mu\nu\,I}$ by a constant
transformation, so that the new linear combinations are all subject to
Bianchi identities. Half of them are then selected as the new field
strengths defined in terms of new gauge fields, while the Bianchi
identities on the remaining linear combinations are regarded as field
equations belonging to a new Lagrangian defined in terms of the new
field strengths. In order that this dualization can be effected the
rotation betweeen the tensors must belong to
$\mathrm{Sp}(2n+2;\mathbb{R})$, where $n+1$ denotes the number of
independent gauge fields. Hence this leads to new quantities $(\tilde
p^I,\tilde q_I)$ and $(\tilde e^I,\tilde f_I)$, where 
\begin{eqnarray}
  \label{eq:em-dual}  
  \tilde p^I &=& U^I{}_J\, p^J + Z^{IJ}\, q_J \,,\nonumber\\
  \tilde q_I &=& V_I{}^J\, q_J + W_{IJ}\, p^J\,, 
\end{eqnarray}
and likewise for $(e^I,f_I)$. Here $U^I{}_J$, $V_I{}^J$, $W_{IJ}$ and
$Z^{IJ}$ are constant real $(n+1)\times(n+1)$ submatrices subject to 
\begin{eqnarray}
  \label{eq:symplectic}
  &&
  U^{\mathrm{T}} V - W^{\mathrm{T}} Z= V^{\mathrm{T}} U
  -Z^{\mathrm{T}} W= \unity\,, \nonumber\\
  &&
  U^{\mathrm{T}}W = W^{\mathrm{T}}U\,,\qquad Z^{\mathrm{T}}V=
  V^{\mathrm{T}}Z \,, 
\end{eqnarray}
so that the full matrix belongs to $\mathrm{Sp}(2n+2;\mathbb{R})$
\cite{Gaillard:1981rj}. Since the charges are not continuous but will
take values in an integer-valued lattice, this group should eventually
be restricted to an appropriate arithmetic subgroup.

Subsequently we define the reduced Lagrangian by the integral of the
full Lagrangian over $S^2$, 
\begin{equation}
  \label{eq:reduced-action}
  \mathcal{F}(e,p,v,u) = \int \mathrm{d}\theta\,\mathrm{d}\varphi\;
  \sqrt{\vert g\vert} \, \mathcal{L} \,.
\end{equation}
We note that the definition of the conjugate quantities $q_I$ and
$f_I$ takes the form, 
\begin{equation}
  \label{eq:q-F-f-F}
  q_I = - \frac{\partial\mathcal{F}}{\partial e^I}\;, \qquad
  f_I =  - \frac{\partial\mathcal{F}}{\partial p^I}\;.
\end{equation}

It is known that a Lagrangian does not transform as a function under
electric/magnetic dualities. Instead we have \cite{deWit:2001pz},
\begin{equation}
  \label{eq:tilde-reduced}
  \tilde{\mathcal{F}}( \tilde e,\tilde p, v,u) +\tfrac1{2}[\tilde
  e^I\tilde q_I + \tilde f_I \tilde p^I ] = 
  \mathcal{F}(e,p, v,u) +\tfrac1{2} [e^I q_I + f_I p^I ] \,. 
\end{equation}
so that the linear combination $\mathcal{F}(e,p, v,u) +\tfrac1{2} [e^I
q_I + f_I p^I]$ transforms as a function. Furthermore one may verify 
that first-order partial derivatives (say with respect to $u$ or $v$,
or derivatives thereof) of $\mathcal{F}(e,p,v,u)$ that leave $e^I$ and
$p^I$ fixed, transform also as a function. This result implies that
the field equations associated with fields other than the
electromagnetic ones transform covariantly and retain their form when
changing the electric/magnetic duality frame.

It is easy to see that the combination $e^I q_I - f_I p^I$ transforms
as a function as well, so that we may construct a modification of
\eqref{eq:reduced-action} that no longer involves the $f_I$ and that 
transforms as a function under electric/magnetic duality,
\begin{equation}
  \label{eq:entropy-function}
  \mathcal{E}(q,p,v,u) =- \mathcal{F}(e,p,v,u) -e^I q_I \,,
\end{equation}
which takes the form of a Legendre transform in view of the first
equation (\ref{eq:q-F-f-F}). In this way we obtain a function of
electric and magnetic charges. Therefore it transforms under
electric/magnetic duality according to $\tilde {\mathcal{E}}(\tilde
q,\tilde p, v,u) = {\mathcal{E}}(q,p, v,u)$. Furthermore the field
equations imply that the $q_I$ are constant and that the action,
$\int\, \mathrm{d}t\mathrm{d}r\, \mathcal{E}$, is stationary under
variations of the fields $v$ and $u$, while keeping the $p^I$ and
$q_I$ fixed. This is to be expected as $\mathcal{E}$ is in fact the
analogue of the Hamiltonian density associated with the reduced
Lagrangian density \eqref{eq:reduced-action}, at least as far as the
vector fields are concerned.

In the near-horizon background (\ref{eq:bar-g}), assuming fields that
are invariant under the $AdS_2$ isometries, the generally covariant
derivatives of the fields vanish and the equations of motion imply
that the constant values of the fields $v_{1,2}$ and $u_{\alpha}$ are
determined by demanding $\mathcal{E}$ to be stationary under
variations of $v$ and $u$,
\begin{eqnarray}
  \label{eq:field-eqs}
  \frac{\partial \mathcal{E}}{\partial v} =  \frac{\partial
  \mathcal{E}}{\partial u} = 0\,,\qquad q_I = \mathrm{constant}\,.
\end{eqnarray}
The function $2\pi\,\mathcal{E}(q, p, v,u)$ coincides with the entropy
function proposed by Sen \cite{Sen:2005wa}. The first two equations of
\eqref{eq:field-eqs} are then interpreted as the attractor equations
and the Wald entropy is directly proportional to the value of
$\mathcal{E}$ at the stationary point,
\begin{equation}
  \label{eq:3}
  \mathcal{S}_{\mathrm{macro}}(p,q) \propto \mathcal{E}
  \Big\vert_{\mathrm{attractor}}\,.  
\end{equation}
The normalization conventions used for the Lagrangian affect $\mathcal{E}$
and the definition of the charges and of Planck's constant.  This has
to be taken into account when determining the proportionality factor
in \eqref{eq:3}, and we do so in \eqref{eq:entropy<function}.  
In the presentation above we followed the approach of
\cite{Sen:2005wa}, but similar approaches can be found in, for
instance, \cite{Ferrara:1997tw,Gibbons:1997cc,Goldstein:2005hq}. Note
that the entropy function does not necessarily depend on all fields at
the horizon. The values of some of the fields will then be left
unconstrained, but those will not appear in the expression for the
Wald entropy.

The above derivation of the entropy function applies to any gauge and
general coordinate invariant Lagrangian, and, in particular, also to
Lagrangians containing higher-derivative interactions.  In the absence
of higher-derivative terms, the reduced Lagrangian $\mathcal{F}$ is at
most quadratic in $e^I$ and $p^I$ and the Legendre transform
\eqref{eq:entropy-function} can easily be carried out. For instance,
consider the following Lagrangian in four space-time dimensions
(we only concentrate on terms quadratic in the field strengths),
\begin{equation}
  \label{eq:quadratic-L}
\sqrt{\vert g\vert} \,\mathcal{L}_{0} =
-\ft14 \, \mathrm{i}\sqrt{\vert g\vert} \Big\{\mathcal{N}_{IJ} 
\, {F}_{\mu\nu}^+{}^{I}\,{F}^{+\mu\nu J}  -
 \bar{\mathcal{N}}_{IJ} \, {F}_{\mu\nu}^-{}^{I}\, 
 {F}^{-\mu\nu J} \Big\} \;,
\end{equation}
where $F^\pm_{\mu\nu}{}^I$ denote the (anti)-selfdual field strengths.
In the context of this paper the tensors
$F^\pm_{\underline{r}\underline{t}}{}^I = \pm \mathrm{i}
F^{\pm}_{\underline{\theta}\underline{\varphi}}{}^I=
\tfrac12(F_{\underline{r}\underline{t}}{}^I \pm \mathrm{i}
F_{\underline{\theta}\underline{\varphi}}{}^I)$ are relevant, where
underlined indices refer to the tangent space.  From
\eqref{eq:quadratic-L}, \eqref{eq:general-fields} and
\eqref{eq:bar-g}, we straightforwardly derive the associated reduced
Lagrangian \eqref{eq:reduced-action},
\begin{equation}
  \label{eq:quadr-reduced-Lagr}
  \mathcal{F} = \tfrac14\left\{ \frac{\mathrm{i} v_1\, p^I (\bar
  {\mathcal{N}} - \mathcal{N})_{IJ}\,  p^J}{4\pi\,v_2} - 
  \frac{4\mathrm{i} \pi\,v_2\, e^I (\bar
  {\mathcal{N}} - \mathcal{N})_{IJ}\, e^J}{v_1}\right\} - \tfrac12 e^I
  (\mathcal{N} + \bar{\mathcal{N}})_{IJ}\, p^J \;. 
\end{equation}
It is straightforward to evaluate the entropy function
\eqref{eq:entropy-function} in this case, 
\begin{equation}
  \label{eq:entropy-quadratic}
  \mathcal{E}= - \frac{v_1}{8\pi\,v_2} \; (q_I - \mathcal{N}_{IK}\,p^K)\,
  [(\mathrm{Im}\,\mathcal{N})^{-1}]^ {IJ}\,  
  (q_J - \bar{\mathcal{N}}_{JL}\,p^L) \;, 
\end{equation}
which is indeed compatible with electric/magnetic duality. Upon
decomposing into real matrices, $\mathrm{i} \mathcal{N}_{IJ}= \mu_{IJ}
- \mathrm{i} \nu_{IJ}$, this result coincides with the corresponding
terms in the so-called black hole potential discussed in
\cite{Ferrara:1997tw,Gibbons:1997cc}, and, more recently, in
\cite{Goldstein:2005hq}. 
\subsection{The BPS entropy function}
\label{sec:bps-entropy-function}
In the previous subsection the symmetry of the near-horizon geometry
played a crucial role. For BPS black holes the supersymmetry
enhancement at the horizon is the crucial input that constrains
certain fields at the horizon as well as the near-horizon geometry.
Unlike in the previous case, the number of attractor equations is
clear and is in principle given by the number of independent
supermultiplets. However, the precise nature of these constraints is
not always a priori clear. For instance, in the case of $N=2$
supergravity, which we will be dealing with in more detail in
subsequent sections, the requirement of supersymmetry enhancement
allows the hypermultiplet scalars to take arbitrary values, while
the value of the vector multiplet scalars is constrained by the black
hole charges.

The $N=2$ vector multiplets contain complex physical scalar fields
which we denote by $X^I$. In supergravity these fields are defined
projectively. At the two-derivative-level, the action for the vector
multiplets is encoded in a holomorphic function $F(X)$. The coupling
to supergravity requires this function to be homogeneous of second
degree. Here we follow the conventions of \cite{LopesCardoso:2000qm},
where the charges and the Lagrangian have different normalizations
than in the previous subsection. However this subsection and the
previous one are self-contained, and the issue of relative
normalizations will only play a role in section \ref{sec:N=2sugra}.
There is one issue, however, that needs to be discussed. In principle
electric/magnetic duality is a feature that pertains to the gauge
fields. Straightforward application of such a duality to an $N=2$
supersymmetric Lagrangian with vector multiplets, leads to a new
Lagrangian that no longer takes the canonical form in terms of a
function $F(X)$. In order to bring it into that form one must
simultaneously apply a field redefinition to the scalar and spinor
fields. On the scalar fields, this redefinition follows from the
observation that $(X^I,F_I(X))$ transforms as a sympletic vector
analogous to the tensors $(F_{\mu\nu}{}^I, G_{\mu\nu I})$ discussed
previously. The need for this field redefinition clearly follows from
the observation that the gauge fields and the fields $X^I$ have a
well-defined relation imposed by supersymmetry. When integrating the
rotated version of the $F_I$ one obtains the new function $\tilde
F(\tilde X)$ in terms of which the new Lagrangian is encoded.
Therefore, in the following, the duality relation of $(X^I,F_I(X))$
will have to be taken into account.  We refer to
\cite{deWit:2001pz,deWit:1996ix} for further details and a convenient
list of formulae.

Upon a suitable uniform field-dependent rescaling of the fields, the
BPS attractor equations take a convenient form\footnote{
  We ignore the hypermultiplets at this stage. } 
which is manifestly consistent with electric/magnetic duality, 
\begin{equation}
  \label{eq:BPS-attractor}
  \mathcal{P}^I=0\,,\qquad \mathcal{Q}_I= 0\,, \qquad \Upsilon=-64\,, 
\end{equation}
where 
\begin{eqnarray}
  \label{eq:P-Q-def}
  \mathcal{P}^I &\equiv& p^I + \mathrm{i}(Y^I-\bar Y^I)  \,,
  \nonumber \\ 
  \mathcal{Q}_I &\equiv& q_I + \mathrm{i}(F_I-\bar F_I)  \,.
\end{eqnarray}
Here the $Y^I$ are related to the $X^I$ by the uniform rescaling and
$F_I$ denotes the derivative of $F(Y)$ with respect to $Y^I$.
Furthermore $\Upsilon$ is a complex scalar field equal to the square
of the $N=2$ auxiliary field $T_{ab}{}^{ij}$ of the Weyl multiplet
(upon the uniform rescaling), which is an anti-selfdual Lorentz
tensor. Note that for fields satisfying the attractor equations
\eqref{eq:BPS-attractor}, one easily establishes that 
\begin{equation}
  \label{eq:Z2}
  \vert Z\vert^2 \equiv  p^I F_I - q_I Y^I \,,
\end{equation}
is equal to $\mathrm{i} (\bar Y^IF_I -Y^I\bar F_I)$ and therefore
real; $Z$ is sometimes refered to as the 'holomorphic BPS mass' and
equals the central charge for the vector supermultiplet system. In
terms of the original variables $X^I$ it is defined as
\begin{equation}
  \label{eq:Z}
    Z = \exp[\mathcal{K}/2] \, (p^I F_I(X) - q_I X^I) \,,
\end{equation}
where 
\begin{equation}
  \label{eq:cal-K}
    \mathrm{e}^{-\mathcal{K}} =  \mathrm{i} \,(\bar X^IF_I(X) - \bar
    F_I(\bar X) X^I)\,. 
\end{equation}
At the horizon the variables $Y^I$ are defined by 
\begin{equation}
  \label{eq:Y-X-horizon}
  Y^I=  \exp[\mathcal{K}/2] \,\bar Z\,X^I \,.
\end{equation}

It is possible to incorporate higher-order derivative interactions
involving the square of the Weyl tensor, by including the Weyl
multiplet into the function $F$, preserving its homogeneity according
to
\begin{equation}
  \label{eq:homogeneous-F}
  F(\lambda Y,\lambda^2 \Upsilon) = \lambda^2\, F( Y,\Upsilon)\,.
\end{equation}
As it turns out \cite{LopesCardoso:2000qm} this modification does not
change the form of the attractor equations \eqref{eq:BPS-attractor}.

The BPS attractor equations can also be described by a variational
principle based on an entropy function
\cite{Behrndt:1996jn,LopesCardoso:2006bg},  
\begin{equation}
  \label{eq:Sigma-simple}
  \Sigma(Y,\bar Y,p,q) =  \mathcal{F}(Y,\bar Y,\Upsilon,\bar\Upsilon)
  - q_I   (Y^I+\bar Y^I) + p^I (F_I+\bar F_I)  \;,
\end{equation}
where $p^I$ and $q_I$ couple to the corresponding magneto- and
electrostatic potentials at the horizon (cf.
\cite{LopesCardoso:2000qm}) in a way that is consistent with
electric/magnetic duality. The quantity $\mathcal{F}(Y,\bar
Y,\Upsilon,\bar\Upsilon)$, which will be denoted as the free energy,
is defined by
\begin{equation}
  \label{eq:free-energy-phase}
  \mathcal{F}(Y,\bar Y,\Upsilon,\bar\Upsilon)= - \im \left( {\bar Y}^I
  F_I - Y^I {\bar F}_I
  \right) - 2\mathrm{i} \left( \Upsilon F_\Upsilon - \bar \Upsilon
  \bar F_{\Upsilon}\right)\,,
\end{equation}
where $F_\Upsilon= \partial F/\partial\Upsilon$. Also this expression
is compatible with electric/magnetic duality \cite{deWit:1996ix}. Varying
the entropy function $\Sigma$ with respect to the $Y^I$, while keeping
the charges and $\Upsilon$ fixed, yields the result,
\begin{equation}
  \label{eq:Sigma-variation-1}
 \delta \Sigma = \mathcal{P}^I
 \, \delta ( F_{I} + \bar F_I)  -\mathcal{Q}_I\,
 \delta (Y^I+ \bar Y^I)  \;.
\end{equation}
Here we made use of the homogeneity of the function $F(Y,\Upsilon)$.
Under the mild assumption that the matrix 
\begin{equation}
  \label{eq:def-N}
  N_{IJ} = \mathrm{i}(\bar F_{IJ}-F_{IJ}), 
\end{equation}
is non-degenerate, it thus follows that stationary points of
$\Sigma$ satisfy the attractor equations. The macroscopic entropy is
equal to the entropy function taken at the attractor point. This
implies that the macroscopic entropy is the Legendre transform of the
free energy $\mathcal{F}$.  An explicit calculation yields the entropy
formula \cite{LopesCardoso:1998wt},
\begin{equation}
  \label{eq:W-entropy}
  {\cal S}_{\rm macro}(p,q) = \pi \, \Sigma\Big\vert_{\rm attractor}
    = \pi \Big[\vert Z\vert^2 - 256\,
  {\rm Im}\, F_\Upsilon \Big ]_{\Upsilon=-64} \,.
\end{equation}
Here the first term represents a quarter of the horizon area (in
Planck units) so that the second term defines the deviation from the
Bekenstein-Hawking area law. In view of the homogeneity properties and
the fact that $\Upsilon$ takes a fixed value the second term will be
subleading in the limit of large charges. Note, however, that also the
area will contain subleading terms, as it will also depend on
$\Upsilon$. In the absence of $\Upsilon$-dependent terms, the
homogeneity of the function $F(Y)$ implies that the area scales
quadratically with the charges.  

We should emphasize that also other higher-derivative interactions can
be present and those will not be captured by the function
$F(Y,\Upsilon)$. We will return to this issue in section
\ref{sec:discussion}.

\section{Application to N=2 supergravity}
\label{sec:N=2sugra}
\setcounter{equation}{0}
We now study the various entropy functions for $N=2$ supergravity
systems. Following \cite{Sahoo:2006rp} we will first determine the
form of the entropy function $\mathcal{E}$. Subsequently we will
exhibit its relation to the BPS entropy function $\Sigma$. The
supergravity Lagrangian consists of various parts. The most important
one concerns the vector multiplets, including the possible effect from
the Weyl multiplet. To this we have to add the Lagrangian for a second
compensating supermultiplet, which we take to be a hypermultiplet.
Other choices for the compensating multiplet (three different choices
have been studied in the literature \cite{deWit:1982na}) are, of
course, possible and should yield identical results. Additional
hypermultiplets may also be added, but play a passive role in the
following. The relevant Lagrangian is given by
\cite{LopesCardoso:2000qm},
\begin{eqnarray}
  \label{eq:efflag}
  8\pi\,e^{-1}\, {\cal L} &=&  
  \im {\cal D}^{\mu} F_I \, {\cal D}_{\mu} \bar X^I   - \im F_I\,\bar X^I 
  (\ft16  R - D) 
  -\ft18\im  F_{IJ}\, Y_{ij}{}^I\, Y^{Jij} - \ft14 \im \hat 
  B_{ij}\,F_{{ A}I}  Y^{Iij}   \nonumber\\
  &&
  +\ft14 \im F_{IJ} (F^{-I}_{ab} -\ft 14 \bar X^I 
  T_{ab}{}^{ij}\varepsilon_{ij})(F^{-Jab} -\ft14 \bar X^J 
  T^{ijab}\varepsilon_{ij})  \nonumber\\
  &&
  -\ft18 \im F_I(F^{+I}_{ab} -\ft14  X^I 
  T_{abij}\varepsilon^{ij}) T^{ab}{}_{ij}\varepsilon^{ij}  
  +\ft12 \im \hat F^{-ab}\, F_{{ A}I} (F^{-I}_{ab} - \ft14  \bar X^I 
  T_{ab}{}^{ij}\varepsilon_{ij})   \nonumber \\
  &&
  +\ft12 \im F_{A} \hat C -\ft18 \im F_{ A A}(\varepsilon^{ik}
  \varepsilon^{jl}  \hat B_{ij} 
  \hat B_{kl} -2 \hat F^-_{ab}\hat F^{-ab}) 
  -\ft1{32} \im F (T_{abij}\varepsilon^{ij})^2 + {\rm h.c.}\nonumber \\ 
  && 
  - \ft12 \varepsilon^{ij}\, \bar \Omega_{\alpha\beta} \, 
  {\cal D}_\mu A_i{}^\alpha \,{\cal D}^\mu  A_j{}^\beta +
  \chi (\ft16 R+  \ft12  D) \;, 
\end{eqnarray}
where the last two terms pertain to the hypermultiplets. This
expression is consistent with electric/magnetic duality upon use of
the field equations for the vector fields and the auxiliary fields
$Y_{ij}{}^I$ \cite{deWit:1996ix}. The quantities $A_i{}^\alpha(\phi)$
denote the hypermultiplet sections, and $\chi$ denotes the
hyper-K\"ahler potential. We refrain from giving explicit definitions
at this point and refer the reader to \cite{dWKV:1999}. The covariant
derivatives involve all the bosonic gauge fields, such as the Lorentz
spin connection and the gauge fields associated with Weyl rescalings
and the $\mathrm{SU}(2)\times\mathrm{U}(1)$ R-symmetry.  The
quantities $X^I$, $F_{ab}^\pm{}^I$ and $Y_{ij}{}^I$ denote the bosonic
components of the vector multiplets, namely, the complex scalars, the
(anti-)selfdual field strengths (defined with tangent-space indices)
and the auxiliary fields, respectively. As we already explained, the
anti-selfdual tensor field $T_{ab}{}^{ij}$ belongs to the Weyl
multiplet and defines the lowest component of a scalar chiral
multiplet, $\hat A=(T_{ab}{}^{ij} \varepsilon_{ij})^2$, which, upon
rescaling yields the field $\Upsilon$ introduced earlier.  Apart from
$T_{ab}{}^{ij}$, the bosonic components of the Weyl multiplet comprise
the Riemann curvature, the field strengths of the
$\mathrm{SU}(2)\times\mathrm{U}(1)$ gauge fields associated with
R-symmetry, and a real scalar field denoted by $D$. The quantities
$\hat B_{ij}$, $\hat F_{ab}^\pm$ and $\hat C$ denote the other bosonic
components of the scalar chiral multiplet constructed from the Weyl
multiplet. For the exact expressions we refer to
\cite{LopesCardoso:2000qm}.

The fact that we extracted a uniform factor of $8\pi$ from the
Lagrangian and the fact that the charges used in
\cite{LopesCardoso:2000qm} differ 
from the charges introduced in \eqref{eq:general-fields} and in 
\eqref{eq:def-q-f}, implies that the charges $p^I$ and $q_I$ as
defined in subsection~\ref{sec:reduced-action} should be changed
according to: $p^I\to 4\pi\,p^I$ and $q_I\to \ft12 q_I$. This
rescaling has been carried out in all subsequent formulae.

The next step is to exploit the spherical symmetry and derive the
reduced Lagrangian \eqref{eq:reduced-action}. For the space-time
metric and the field strengths this was already done in
\eqref{eq:general-fields}. Let us first concentrate on the auxiliary
field $T_{ab}{}^{ij}$, which plays an important role in this paper. In
a spherically symmetric configuration this field can be expressed in
terms of a single complex scalar $w$. Following \cite{Sahoo:2006rp} we
define,
\begin{equation}
  \label{eq:T-w}
  T_{\underline{r}\underline{t}}{}^{ij} \varepsilon_{ij} =- \im\, 
  T_{\underline{\theta}\underline{\varphi}}{}^{ij} \varepsilon_{ij} =
  w\,, 
\end{equation}
where underlined indices denote tangent-space indices. Consequently we
have $\hat A= -4 w^2$. We will have to do the same for all other
fields, but we will restrict ourselves to a restricted class of
solutions by putting some of the fields to zero. Namely, at this stage we
will assume the following consistent set of constraints,
\begin{eqnarray}
  \label{eq:restricted-solution}
  R(\mathcal{V})_{\mu\nu}{}^i{}_j =  R({A})_{\mu\nu} =
  \mathcal{D}_\mu X^I =  \mathcal{D}_\mu A_i{}^\alpha =  0 \,,  
\end{eqnarray}
where the first two tensors denote the R-symmetry field strengths.
These constraints are weaker than the ones imposed in
\cite{Sahoo:2006rp}, and they are in accord with those that follow
from requiring supersymmetry enhancement at the horizon
\cite{LopesCardoso:2000qm}. It is not unlikely that, if one were to
relax these constraints in the evaluation of the reduced Lagrangian,
most of them would still emerge in the form of attractor equations at
the end. We will not pursue this question in any detail.

Since $\hat B_{ij}$ is proportional to
$R(\mathcal{V})_{\mu\nu}{}^i{}_j$, this field can thus be ignored as
well. Furthermore the auxiliary fields $Y_{ij}{}^I$ can be dropped
as a result of their equations of motion. Subject to all these
conditions the relevant expressions for $\hat C$ and $\hat F_{\mu\nu}$
are as follows,
\begin{eqnarray}
  \label{eq:hat-F-C}
  \hat F^{-ab}  &=& -16 \,{\cal R}(M)_{cd}{}^{\!ab} \,
T^{klcd}\,\varepsilon_{kl}   \,,\nonumber\\
\hat C &=&  64\, {\cal R}(M)^{-cd}{}_{\!ab}\, {\cal 
R}(M)^-_{cd}{}^{\!ab}  - 32\, T^{ab\,ij} \, D_a \,D^cT_{cb\,ij}  \,,
\end{eqnarray}
where $\mathcal{R}$ is a modification of the Riemann tensor and the
derivatives are superconformally invariant \cite{LopesCardoso:2000qm}.
Under the same assumptions the Lagrangian \eqref{eq:efflag} reduces to
$\mathcal{L} =\mathcal{L}_1+\mathcal{L}_2$, with
\begin{eqnarray}
  \label{eq:efflag-1-2}
  8\pi\,e^{-1}\, {\cal L}_1 &=&
    \Big[\ft14 \im F_{IJ} F^{-I}_{ab} 
  (F^{-Jab} -\ft12 \bar X^J T^{ijab}\varepsilon_{ij})  \nonumber\\
  &&\quad
  -\ft18 \im F_I\,F^{+I}_{ab}
  \, T^{ab}{}_{ij}\varepsilon^{ij}  
  + \tfrac12\im \hat F^{-ab}\, F_{{ A}I} \,F^{-I}_{ab} 
  + {\rm h.c.}\Big] \;,  \nonumber \\[1mm]
   8\pi\,e^{-1}\, {\cal L}_2 &=&   \mathrm{e}^{-\mathcal{K}} 
  (D- \ft16 R) +\ft12 \chi (D + \ft13 R)  
   \nonumber\\
  &&
  -\ft1{32} \Big[\im (F-F_I X^I + \tfrac12 \bar F_{IJ} X^IX^J)
  (T_{abij}\varepsilon^{ij})^2 + {\rm h.c.}\Big] 
  \nonumber\\
  &&
  +\ft12 \Big[ \im F_{A} \hat C +\ft12 \im F_{ A A}\, \hat
  F^-_{ab}\hat F^{-ab}   -\tfrac14\mathrm{i} \hat F^{-ab}\, F_{{
  A}I}\bar X^I  T_{ab}{}^{ij}\varepsilon_{ij}
  + {\rm h.c.}\Big]  \;.  
\end{eqnarray}

In the $AdS_2$ background we are left with a
restricted number of field variables that are all constant, namely,
$v_1$, $v_2$, $w$, $D$, $e^I$, $X^I$ and $A_i{}^\alpha$. Note,
however, that the dependence on the fields $A_i{}^\alpha$ is entirely
contained in the hyperk\"ahler potential $\chi$. Our next task 
is to evaluate the reduced Lagrangian as a function of these
variables. Before doing so, we should stress that the above Lagrangian
\eqref{eq:efflag} was derived from a superconformally invariant
expression. As a result the bosonic quantities are still
subject to certain invariance transformations. One of them is scale
invariance with respect to a complex parameter $\lambda$,
\begin{eqnarray}
  \label{eq:scale-invariance}
  v_{1,2} \to \vert\lambda\vert^{-2} v_{1,2} \,,\quad
  w \to \bar \lambda  w \,,\quad
  D \to \vert\lambda\vert^{2} D \,,\quad
  X^I \to \bar \lambda X^I \,,\quad 
  \chi\to \vert \lambda\vert^2 \chi \,.
\end{eqnarray}
All other fields (as well as the charges) are invariant under these
scale transformation. In addition the hypermultiplet sections are
subject to rigid $\mathrm{SU}(2)$ transformations. The reduced
Lagrangian and the entropy function should be invariant under these
transformations.  Therefore it will be useful to express the entropy
function \eqref{eq:entropy-function} computed from the Lagrangian
\eqref{eq:efflag} in terms of a set of scale invariant variables.  We
choose the following set of such variables, 
\begin{eqnarray}
  \label{eq:rescaledX}
  && 
  Y^I = \ft14 v_2 \, {\bar w} \, X^I \,,\quad
  \Upsilon = \ft{1}{16} v_2^2 \, {\bar w}^2 \, {\hat A} = - \ft14
  v_2^2 \, \vert w \vert^4 \,, \quad 
  U = \frac{v_1}{v_2} \,,\nonumber\\ 
  &&
  {\tilde D} = v_2 \, D + \tfrac23(U^{-1} -1) \,,\quad 
  \tilde\chi=  v_2 \, \chi \;.  
\end{eqnarray}
Observe that $\Upsilon$ is real and negative, and that
$\sqrt{-\Upsilon}$ and $U$ are real and positive. Note also that the
hypermultiplets contribute only through the hyperk\"ahler potential
$\chi$. 

We now compute the quantities appearing in \eqref{eq:efflag-1-2} for
the near-horizon background specified in terms of the parameters given
above. We obtain (indices $i,j$ refer to the $AdS_2$ coordinates
$r,t$, whereas indices $\alpha,\beta$ refer to $S^2$ coordinates
$\theta,\varphi$),
\begin{eqnarray}
  \label{eq:background-quantities}
  R &=& 2 \left( v_1^{-1} - v_2^{-1} \right) \;,\nonumber\\
  f_i{}^j &=& [\ft12 v_1^{-1} - \ft14 (D + \ft13 R) -\ft{1}{32}  
    \vert w \vert^2]\,\delta_i{}^j \;, \nonumber\\
  f_\alpha{}^\beta &=& [ -\ft12 v_2^{-1} - \ft14 (D + \ft13 R)
  +\ft{1}{32} \vert w \vert^2 ]\,\delta_\alpha{}^\beta \;, \nonumber\\
  \mathcal{R}(M)_{ij}{}^{kl} &=& (D+\ft13 R)\,\delta_{ij}{}^{kl} \;,
  \nonumber \\  
  \mathcal{R}(M)_{\alpha\beta}{}^{\gamma\delta} &=& (D+\ft13
  R)\,\delta_{\alpha\beta}{}^{\gamma\delta} \;,   \nonumber \\    
   \mathcal{R}(M)_{i\alpha}{}^{j \beta} &=& \ft12 (D- \ft16
  R)\,\delta_i^j\,\delta_\alpha^\beta \;,   \nonumber \\    
  {\hat A} &=&  - 4 w^2 \;, \nonumber\\
  {\hat F}^-_{\underline{r}\underline{t}} &=&- 
  \mathrm{i} {\hat F}^-_{\underline{\theta}\underline{\varphi}}=- 16 w (D
  + \ft13 R) \;,   \nonumber\\ 
  {\hat C} &=& 192 D^2 +   \ft{32}3  R^2 
  - 16 \vert w \vert^2 ( v_1^{-1}+ v_2^{-1}) +2\vert w \vert^4 \;.
\end{eqnarray} 
With these results we obtain the following contributions to the
reduced Lagrangian corresponding to $\mathcal{L}_1$ and
$\mathcal{L}_2$ of \eqref{eq:efflag-1-2}, 
\begin{eqnarray}
  \label{eq:F1-F2}
  \mathcal{F}_1 &=&{}  \tfrac18 N_{IJ} \Big[  U^{-1}
  e^Ie^J - U {p^I} {p^J} \Big] -
  \tfrac14(F_{IJ}+ \bar F_{IJ}) {e^Ip^J}  \nonumber \\
  &&
  + \ft12 \mathrm{i} e^I\Big[ F_I + F_{IJ} \bar Y^J + 8 
  F_{I\Upsilon}\sqrt{-\Upsilon}\tilde D  - \mathrm{h.c.}\Big] \nonumber\\
  &&
  - \ft12 U p^I\Big[ F_I - F_{IJ} \bar Y^J - 8  
  F_{I\Upsilon}\sqrt{-\Upsilon}\tilde D +  \mathrm{h.c.}\Big]\;,
  \nonumber\\[2mm] 
  \mathcal{F}_2 &=&  
  \frac{4\mathrm{i}}{\sqrt{-\Upsilon}} (\bar Y^IF_I -Y^I\bar F_I)
  (\tilde D U +U -1) + \ft14 \tilde \chi\tilde DU \nonumber\\
  &&
  + \mathrm{i} U \Big[ F-Y^IF_I - 2\Upsilon F_\Upsilon+  \tfrac12 \bar
  F_{IJ}Y^IY^J - \mathrm{h.c.} \Big] \nonumber\\
  &&
  + \mathrm{i}(F_\Upsilon-\bar F_\Upsilon) \Big [48 U \tilde D^2 +64
  \tilde D(U-1) + 32 (U+U^{-1} -2) - 8 (1+U) \sqrt{-\Upsilon} \Big]
  \nonumber\\ 
  &&
  +32 \mathrm{i}U \Big[\tilde D^2 \Upsilon F_{\Upsilon\Upsilon} -
  \tfrac14 \tilde D \,\bar Y^I F_{I\Upsilon} \sqrt{-\Upsilon} -
  \mathrm{h.c.} \Big] \;.  
\end{eqnarray}
Observe that these results refer to a general function
$F(Y,\Upsilon)$. Because of the scale invariance, there is no longer a
dependence on the field $w$. Furthermore we used the definition
\eqref{eq:def-N}. 

The entropy function can be written as 
\begin{equation}
  \label{eq:entropy-1+2}
  \mathcal{E} = \mathcal{E}_1 + \mathcal{E}_2\,, 
\end{equation}
where $\mathcal{E}_1= -\mathcal{F}_1 - \ft12 e^Iq_I$ and
$\mathcal{E}_2= -\mathcal{F}_2$. Note that the factor $1/2$ in
$\mathcal{E}_1$ is due to the rescaling discussed earlier. When expressed
in terms of $p^I$ and $q_I$, $\mathcal{E}_1$ reads, 
\begin{eqnarray}
  \label{eq:entropy-1}
   \mathcal{E}_1  &=& 
   \ft12 U\,\Sigma(Y, \bar Y,p,q) +\ft12 U\,N^{IJ} (\mathcal{Q}_I-F_{IK}
   \mathcal{P}^K)\, (\mathcal{Q}_J-\bar F_{JL} \mathcal{P}^L) 
   \nonumber\\
   &&
   + \mathrm{i}U \Big[ \Upsilon F_\Upsilon-\ft12 Y^I F_I + \ft12 \bar 
   F_{IJ}Y^I Y^J  - 
   \mathrm{h.c.}\Big] \nonumber\\
   &&
   +8\mathrm{i} U\tilde D \sqrt{-\Upsilon} \Big[F_{I\Upsilon} N^{IJ}
   (\mathcal{Q}_J - \bar F_{JK}\mathcal{P}^K) - \mathrm{h.c.}\Big]
   \nonumber\\ 
   &&
   -8\mathrm{i} U\tilde D \sqrt{-\Upsilon} \Big[\bar Y^I F_{I\Upsilon}
   - \mathrm{h.c.}\Big]
   \nonumber\\ 
   &&
   + 32 U\,  \tilde D^2 \Upsilon \,N^{IJ} (F_{I\Upsilon}-\bar
   F_{I\Upsilon}) (F_{J\Upsilon}-\bar F_{J\Upsilon}) \,,
\end{eqnarray}
where $\mathcal{Q}_I$, $\mathcal{P}^I$, and $\Sigma$ were defined
already in \eqref{eq:P-Q-def} and \eqref{eq:Sigma-simple},
respectively. Combining this result with $\mathcal{E}_2$ there are
some crucial rearrangements and the result is an entropy function that
is consistent with electric/magnetic duality,
\begin{eqnarray}
  \label{eq:entropy-total-1}
   \mathcal{E}  &=& 
   \ft12 U\,\Sigma(Y, \bar Y,p,q) +\ft12 U\,N^{IJ} (\mathcal{Q}_I-F_{IK}
   \mathcal{P}^K)\, (\mathcal{Q}_J-\bar F_{JL} \mathcal{P}^L) 
   \nonumber\\
   &&
   +8\mathrm{i} U\tilde D \sqrt{-\Upsilon} \Big[F_{I\Upsilon} N^{IJ}
   (\mathcal{Q}_J - \bar F_{JK}\mathcal{P}^K) - \mathrm{h.c.}\Big]
   \nonumber\\ 
   &&
   -\frac{4\mathrm{i}}{\sqrt{-\Upsilon}} (\bar Y^IF_I -Y^I\bar F_I)
   (\tilde D U +U -1) - \ft14 \tilde \chi\tilde DU \nonumber\\
  &&
  -32 \mathrm{i}U \tilde D^2\Big[\Upsilon F_{\Upsilon\Upsilon} +\ft12 
   \mathrm{i}  \Upsilon \,N^{IJ} (F_{I\Upsilon}-\bar
   F_{I\Upsilon}) (F_{J\Upsilon}-\bar F_{J\Upsilon}) 
   - \mathrm{h.c.} \Big] \nonumber\\
  &&
  - \mathrm{i}(F_\Upsilon-\bar F_\Upsilon) \Big [48 U \tilde D^2 +64
  \tilde D(U-1) - 2 U\Upsilon + 32 (U+U^{-1} -2) - 8 (1+U)
   \sqrt{-\Upsilon} \Big] \;,
  \nonumber\\ 
  &&
\end{eqnarray}
where we used the homogeneity of the function $F(Y,\Upsilon)$, which
implies 
\begin{equation}
  \label{eq:homo-F}
   F(Y,\Upsilon) = \ft12 Y^I F_I(Y,\Upsilon) + \Upsilon
  F_\Upsilon(Y,\Upsilon)\,.  
\end{equation}
To confirm that the entropy transforms as a function under
electric-magnetic duality, one may make use of the results of
\cite{deWit:1996ix}.  Subsequently we require that $\mathcal{E}$ be
stationary with respect to variations of $\tilde D$ and $\tilde\chi$.
This imposes the conditions (we assume $U\neq 0$),
\begin{eqnarray}
  \label{eq:tilde-D-chi}
  \tilde D &=& 0\,, \nonumber \\
  \tilde \chi &=& -\frac{16\mathrm{i}}{\sqrt{-\Upsilon}} (\bar Y^IF_I
  -Y^I\bar F_I) - 256 \mathrm{i} (F_\Upsilon-\bar F_\Upsilon)
  (1-U^{-1}) \nonumber \\
  &&
  + 32 \mathrm{i} \sqrt{-\Upsilon}
  \Big[F_{I\Upsilon} N^{IJ} 
   (\mathcal{Q}_J - \bar F_{JK}\mathcal{P}^K) - \mathrm{h.c.}\Big] \,.
\end{eqnarray}
Upon substitution of these equations into \eqref{eq:entropy-total-1},
the expression for $\mathcal{E}$ simplifies considerably and we
obtain,
\begin{eqnarray}
  \label{eq:entropy-total-2}
   \mathcal{E}(Y,\bar Y,\Upsilon,U)  &=& 
   \ft12 U\,\Sigma(Y, \bar Y,p,q) +\ft12 U\,N^{IJ} (\mathcal{Q}_I-F_{IK}
   \mathcal{P}^K)\, (\mathcal{Q}_J-\bar F_{JL} \mathcal{P}^L) 
   \nonumber\\
   &&
   -\frac{4\mathrm{i}}{\sqrt{-\Upsilon}} (\bar Y^IF_I -Y^I\bar F_I)
   (U -1)   \nonumber\\
  &&
  - \mathrm{i}(F_\Upsilon-\bar F_\Upsilon) \Big [- 2 U\Upsilon + 32
   (U+U^{-1} -2) - 8 (1+U) \sqrt{-\Upsilon} \Big] \;. 
\end{eqnarray}
Although this result is written in a different form and is obtained in
a slightly different setting, it is in accord with the result derived
in \cite{Sahoo:2006rp}.  The entropy function
\eqref{eq:entropy-total-2} depends on the variables $U$, $\Upsilon$
and $Y^ I$ whose values will be determined at the attractor values
where $\mathcal{E}$ is stationary. The macroscopic entropy is
proportional to the entropy function taken at the attractor values,
\begin{equation}
  \label{eq:entropy<function}
  \mathcal{S}_{\mathrm{macro}}(p,q) = 2\pi
  \mathcal{E}\Big\vert_{\mathrm{attractor}} \,. 
\end{equation}
In the following, we will discuss the extremization of $\cal E$ with
respect to these variables, first in the absence of $R^2$-terms, and
then for BPS black holes in the presence of $R^2$-terms. Finally we
will consider the general case. 

\subsection{Variational equations without $R^2$-interactions}
\label{sec:variation-without-R2}
In the absence of $R^2$-interactions, the function $F$ does not depend
on $\Upsilon$, so that the entropy function \eqref{eq:entropy-total-2}
reduces to
\begin{eqnarray}
  \label{eq:entropy-total-3}
   \mathcal{E}(Y,\bar Y,\Upsilon,U)  &=& 
   \ft12 U\,\Sigma(Y, \bar Y,p,q) +\ft12 U\,N^{IJ} (\mathcal{Q}_I-F_{IK}
   \mathcal{P}^K)\, (\mathcal{Q}_J-\bar F_{JL} \mathcal{P}^L) 
   \nonumber\\
   &&
   -\frac{4\mathrm{i}}{\sqrt{-\Upsilon}} (\bar Y^IF_I -Y^I\bar F_I)
   (U -1)   \,.
\end{eqnarray}
Varying \eqref{eq:entropy-total-3} with respect to $\Upsilon$ yields
\begin{equation}
  \label{eq:eq-Upsilon}
  U = 1 \,.
\end{equation}
The latter implies that the Ricci scalar of the four-dimensional
space-time vanishes. Here we assumed that $\left(\bar Y^I F_I - Y^I
  \bar F_I \right )$ is non-vanishing, which is required so that
Newton's constant remains finite. Varying with respect to $U$ yields,
\begin{eqnarray}
  \label{eq:eq-U}
  \Sigma +  \left( {\cal Q}_I - F_{IK} \, {\cal P}^K\right) N^{IJ} 
  \left( {\cal Q}_J - {\bar F}_{JL} \, {\cal P}^L \right)
  - \frac{8\mathrm{i}}{\sqrt{-\Upsilon}} 
  \left(\bar Y^I F_I -Y^I \bar F_I \right ) =0 \;,
\end{eqnarray}
which determines the value of $\Upsilon$ in terms of the $Y^I$. This
relation is not surprising. When the function $F$ depends exclusively
on the $Y^I$, the quantity $\Upsilon$ is related to an auxiliary field
in the original Lagrangian whose field equation is algebraic and
\eqref{eq:eq-U} is a direct consequence of this equation.

Hence we are now dealing with an effective entropy function 
\begin{equation}
  \label{eq:effective-entropy-function}
  \mathcal{E}(Y,\bar Y,\Upsilon,1) = \ft12 \Sigma(Y, \bar Y,p,q)
   +\ft12 N^{IJ} (\mathcal{Q}_I-F_{IK} 
   \mathcal{P}^K)\, (\mathcal{Q}_J-\bar F_{JL} \mathcal{P}^L)\,,
\end{equation}
which is independent of $\Upsilon$, whose value is simply determined
by \eqref{eq:eq-U}. Note that \eqref{eq:effective-entropy-function} is
homogeneous under uniform rescalings of the charges $q_I$ and $p^I$
and the variables $Y^I$. This implies that the entropy will be
proportional the the square of the charges. Under infinitesimal
changes of $Y^I$ and $\bar Y^I$ the entropy function
\eqref{eq:effective-entropy-function} changes according to
\begin{eqnarray}
  \label{eq:attractorytree}
  \delta\mathcal{E} &=& 
  \mathcal{P}^I \,\delta(F_I+\bar F_I) -\mathcal{Q}_I \,\delta(Y^I+\bar
  Y^I) \nonumber\\
  && 
  + \tfrac1{2}\mathrm{i} \left({\cal Q}_K -\bar F_{KM} \,{\cal P}^M
  \right) 
  N^{KI} \,\delta F_{IJ} \, N^{JL} \left( {\cal Q}_L - {\bar F}_{LN}
  \,{\cal P}^N \right)\nonumber \\
  && 
  - \tfrac1{2}\mathrm{i} \left({\cal Q}_K -F_{KM} \,{\cal P}^M
  \right) 
  N^{KI} \,\delta \bar F_{IJ} \, N^{JL} \left( {\cal Q}_L -  F_{LN}
  \, {\cal P}^N \right) = 0 \;, 
\end{eqnarray}
where $\delta F_I= F_{IJ}\,\delta Y^J$ and
$\delta{F}_{IJ}=F_{IJK}\,\delta{Y}^K$. This equation determines the
horizon value of the $Y^I$ in terms of the black hole charges $(p^I,
q_I)$. Because the function $F(Y)$ is homogeneous of second degree, we
have $F_{IJK}Y^K=0$. Using this relation one deduces from
\eqref{eq:attractorytree} that $\left({\cal Q}_J - F_{JK} \, {\cal
    P}^K \right) Y^J = 0$, which is equivalent to
\begin{equation}
  \label{eq:Z-Kappa}
  \mathrm{i} ({\bar Y}^I F_I - Y^I{\bar F}_I) = p^I F_I - q_I Y^I \;. 
\end{equation}
Therefore, at the attractor point, we have 
\begin{equation}
\label{eq:expsig}
  \Sigma = \mathrm{i} ( {\bar Y}^I F_I - Y^I {\bar F}_I ) \;.
\end{equation}
Inserting this result into (\ref{eq:eq-U}) yields
\begin{equation}
  \label{eq:valueups}
  \sqrt{- \Upsilon}  = \frac{8\,\Sigma}{\Sigma + N^{IJ} \left(
        {\cal Q}_I - F_{IK} \, {\cal P}^K\right) 
      \left( {\cal Q}_J - {\bar F}_{JL} \, {\cal P}^L \right)} \;,
\end{equation}
which gives the value of $\Upsilon$ in terms of the attractor values
of the $Y^I$. Using (\ref{eq:valueups}) we can write the entropy as, 
\begin{eqnarray}
  \label{eq:entrotree}
  {\cal S}_{\rm macro} (p,q) = 2 \pi \, 
  {\cal E} \Big\vert_{\rm attractor} = 
  \frac{8 \pi\, \Sigma}{\sqrt{-\Upsilon}}\Big\vert_{\rm attractor}  \;.
\end{eqnarray}
Observe that, for a BPS black hole, ${\cal Q}_I = {\cal P}^J =0$ and
$\Upsilon = - 64$, so that ${\cal S}_{\rm macro} = \pi \,
\Sigma\vert_{\rm attractor}$ in accord with \eqref{eq:W-entropy}. 

The entropy function \eqref{eq:effective-entropy-function}
can be written as
\begin{equation}
  \label{eq:Epsilon-U=1} 
  \mathcal{E} = - q_I (Y^I+ {\bar Y}^I) + p^I (F_I +{\bar F}_I )  +
  \ft12 N^{IJ} (q_I-F_{IK}p^K) (q_J-\bar F_{JL}p^L)+ N_{IJ} Y^I\bar
  Y^J  \,,
\end{equation}
where we used the homogeneity of the function $F(Y)$. Expressing the
$Y^I$ according to \eqref{eq:Y-X-horizon} (which is consistent with
the first equation of \eqref{eq:rescaledX}, as we will show below) and
using the definitions \eqref{eq:Z} and \eqref{eq:cal-K}, we write 
\eqref{eq:Epsilon-U=1} as follows,
\begin{equation}
  \label{eq:Epsilon-U=1-X} 
  \mathcal{E} =   \ft12 \left( N^{IJ}+ 2\, \mathrm{e}^{\mathcal{K}} X^I
  \bar X^J\right)  (q_I-F_{IK}p^K) (q_J-\bar F_{JL}p^L)\,, 
\end{equation}
where $F_{IJ}$ is now the second derivative of $F(X)$ with respect to
$X^I$ and $X^J$. Notice that this expression is invariant under
uniform rescalings of the $X^I$ by a complex number, which is a
reflection of the complex scale invariance noted above
\eqref{eq:scale-invariance}. 

The quantities $X^I$ can now be expressed in terms of the physical
complex scalars belonging to the vector supermultiplets, which we
denote by $z^A$, where the index $A$ takes $n$ values, one less than
the number of vector fields. These scalars parametrize the special
K\"ahler target space. Subsequently we parametrize the $X^I$ as a
projective holomorphic section (i.e. up to multiplication by a
complex factor) in terms of the holomorphic coordinates $z^A$. We then
use the identity (see the second reference in \cite{deWit:1996ix}), 
\begin{equation}
  \label{eq:N-inverse-hatX}
  N^{IJ} = {\rm e}^{\mathcal{K}(z,\bar z)} \left[g^{A {\bar B}} 
    \left(\partial_{A} + \partial_{A} \mathcal{K}(z,\bar z)\right)
    X^I(z)\,
    \left(\partial_{\bar B}+\partial_{\bar B} \mathcal{K}(z,\bar z)
    \right) {\bar X}^J (\bar z) - X^I(z) \, {\bar X}^J(\bar z) \right]\;,  
\end{equation}
where $g^{A\bar B}$ is the inverse metric of the special K\"ahler
space, and write the entropy function \eqref{eq:Epsilon-U=1-X} in the
well-known form \cite{Ferrara:1996dd,Ferrara:1997tw},
\begin{equation}
  \label{eq:entropy-hatX}
  \mathcal{E} = \ft12 \left[ \vert Z(z,\bar z) \vert^2  + g^{A {\bar
    B}}(z,\bar z)  \,
    \mathcal{D}_A  Z(z,\bar z)  \,\mathcal{D}_{\bar B} {\bar Z}(z,\bar
    z) \right] \;,
\end{equation}
where $\mathcal{D}_A Z =(\partial_{A}
+\ft12\partial_{A}\mathcal{K})Z$. 
This agreement was also established in
\cite{Cardoso:2006cb}. As mentioned above, in order to bring the
entropy function into the form 
\eqref{eq:entropy-hatX}, we expressed the $Y^I$ according to 
\eqref{eq:Y-X-horizon}, which is consistent with the definition given in 
\eqref{eq:rescaledX} by virtue of~\eqref{eq:Z-Kappa}.

\subsection{BPS black holes with $R^2$-interactions}
\label{sec:bps-black-holes}
In the presence of $R^2$ interactions, the horizon values of $U$ and
$\Upsilon$ for extremal BPS black holes are $U =1$ and $\Upsilon =
-64$ \cite{LopesCardoso:2000qm}.  Inserting these values into
\eqref{eq:entropy-total-2} results in
\begin{eqnarray} 
  \label{eq:entropy-total-2-BPS}
  {\cal E} (Y, {\bar Y}, -64 ,1) 
  =  \ft12  \Sigma (Y, {\bar Y}, p, q) 
  + \ft12 N^{IJ} \left( {\cal Q}_I - F_{IK} \, {\cal P}^K\right) 
  \left( {\cal Q}_J - {\bar F}_{JL} \, {\cal P}^L \right) \;.
\end{eqnarray}
Observe that the variational principle based on 
\eqref{eq:entropy-total-2-BPS} is only consistent
with the one based on \eqref{eq:entropy-total-2}
provided that 
\eqref{eq:entropy-total-2-BPS} is supplemented by
the extremization equations for $U$ and for $\Upsilon$ given by 
\eqref{eq:valueU} and \eqref{eq:homo-delta}
below.  For BPS solutions it can be readily checked
that the latter are indeed satisfied.

The form of the BPS entropy function \eqref{eq:entropy-total-2-BPS}
is closely related to the one
given in \cite{Behrndt:1996jn,LopesCardoso:2006bg}, which consists of
the first term in \eqref{eq:entropy-total-2-BPS}.  As discussed in
section \ref{sec:bps-entropy-function}, the BPS attractor equations
can be derived by a variational principle based on $\Sigma$.  The
quantity $\Sigma$ was also used in \cite{LopesCardoso:2006bg} to
construct a duality invariant version of the OSV integral which
attempts to express microscopic state degeneracies in terms of
macroscopic data \cite{Ooguri:2004zv}.  In \cite{LopesCardoso:2006bg}
it was furthermore shown that, for large BPS black holes, the
evaluation in saddle-point approximation of the modified OSV integral
precisely yields the macroscopic entropy \eqref{eq:W-entropy}.  This
result was established by computing the second variation of $\Sigma$
which, upon imposing the BPS attractor equations ${\cal Q}_I = {\cal
  P}^J =0$, equals $\delta^2 \Sigma = 2 N_{IJ} \, \delta Y^I \, \delta
{\bar Y}^J$.  Instead of constructing a duality invariant version of
the OSV integral based on $\Sigma$, one can also consider constructing
such an integral based on \eqref{eq:entropy-total-2-BPS}.  The
presence of the second term in \eqref{eq:entropy-total-2-BPS} will,
however, not affect the evaluation of this integral in saddle-point
approximation (for large black holes), since, when evaluating the
second variation of ${\cal E}$ on the BPS attractor, the second term
contributes the same amount as the first term, so that $\delta^2 {\cal
  E} =\delta^2 \Sigma = 2 N_{IJ} \, \delta Y^I \, \delta {\bar Y}^J$.

\subsection{Non-BPS black holes with $R^2$-interactions}
\label{sec:non-bps-black-holes}
In the following, we consider extremal black holes in the
presence of  $R^2$-terms and we compute the extremization
equations for the fields $U, \Upsilon$ and $Y^I$ following from 
the entropy function \eqref{eq:entropy-total-2}.

Varying with respect to $U$ gives
\begin{eqnarray}
  \label{eq:valueU} 
  && \Sigma + \left( {\cal Q}_I - F_{IK} \, {\cal P}^K\right) 
  N^{IJ} \left({\cal Q}_J -{\bar F}_{JL} \, {\cal P}^L\right)
   -\frac{8\mathrm{i}}{\sqrt{-\Upsilon}} (\bar Y^IF_I -Y^I\bar F_I)
     \nonumber\\
  &&
  - \mathrm{i}(F_\Upsilon-\bar F_\Upsilon) \Big [- 4 \Upsilon + 64
   (1-U^{-2}) - 16  \sqrt{-\Upsilon} \Big] =0 \;. 
\end{eqnarray}
To verify the consistency with the analysis of the previous subsection
(see \eqref{eq:entropy-total-2-BPS}), one may verify that the BPS
conditions $\mathcal{P}^I = \mathcal{Q}_I=0$ and $\Upsilon= -64$
leaves only the term proportional to $(1-U^{-2}) (F_\Upsilon-\bar
F_\Upsilon)$ which vanishes as a result of $U=1$.

Subsequently we consider the variation of the entropy function
\eqref{eq:entropy-total-2} with respect to arbitrary variations of the
fields $Y^I$ and $\Upsilon$ and their complex conjugates. Denoting
this variation by $\delta= \delta Y^I\partial/\partial Y^I+ \delta\bar
Y^I\partial/\partial \bar Y^I+ \delta\Upsilon\partial/\partial
\Upsilon+ \delta\bar \Upsilon\partial/\partial \bar\Upsilon$, we
derive the following result, 
\begin{eqnarray}
  \label{eq:deltaentro}  
  \delta {\cal E} &=& U \left[
   \mathcal{P}^I \,\delta(F_I+\bar F_I) -  \mathcal{Q}_I \,\delta(Y^I+\bar
  Y^I) \right] \nonumber\\
  &&  
  +\tfrac1{2} \mathrm{i}U \left[({\cal Q}_K -\bar F_{KM} \,{\cal P}^M)
  N^{KI} \,\delta F_{IJ} \, N^{JL} ( {\cal Q}_L - {\bar F}_{LN}
  \,{\cal P}^N )   - \mathrm{h.c.} \right] 
  \nonumber \\
  && 
  - 4 \mathrm{i} (- \Upsilon)^{-1/2}\, (U-1) \,
  \left[ (F_I - {\bar F}_I) \,\delta(Y^I + {\bar Y}^I) 
    - (Y^I - {\bar Y}^I) \, \delta (F_I + {\bar F}_I) \right]
  \nonumber\\
  &&
  + \mathrm{i} \left[ 2 U \, \Upsilon - 32 (U + U^{-1} -2) 
    +16 \sqrt{-\Upsilon} \right] 
  \delta (F_{\Upsilon} - {\bar F}_{\Upsilon}) \nonumber\\
  && 
  + \mathrm{i} U  \left[\delta\Upsilon\,F_{\Upsilon I} N^{IJ} ( {\cal
   Q}_J - {\bar F}_{JL} \,{\cal P}^L) - \mathrm{h.c.} \right] 
   \nonumber\\
   &&- 2 \mathrm{i} (-\Upsilon)^{-3/2} \, (U-1) \, 
   ({\bar Y}^I F_I - Y^I {\bar F}_I )
   \, \delta \Upsilon \nonumber\\
   &&+  \mathrm{i}  (F_{\Upsilon} - {\bar F}_{\Upsilon}) \left[ U - 
     4(-\Upsilon)^{-1/2} \, (1 + U) \right] \, \delta \Upsilon \;, 
\end{eqnarray}
where we took into account that the variable $\Upsilon$ is real.

Restricting ourselves to variations $\delta Y^I$, the above result
leads to the following attractor equations, 
\begin{eqnarray}
  \label{eq:variational-y}
  &&
  U \left(\mathcal{Q}_I - F_{IJ} \, \mathcal{P}^J\right) 
   -\tfrac1{2} \mathrm{i}U \left({\cal Q}_K -\bar F_{KM} \,{\cal P}^M
  \right) 
  N^{KP} \,F_{PIQ} \, N^{QL} \left({\cal Q}_L -{\bar F}_{LN}\,{\cal
  P}^N \right) \nonumber\\
  && 
  + 4 \mathrm{i}(- \Upsilon)^{-1/2} (U-1) 
  \left[F_I - {\bar F}_I - F_{IJ} (Y^J -{\bar Y}^J) \right]
  \nonumber\\ 
  &&
  -  \mathrm{i} \left[ 2 U \, \Upsilon - 32 (U + U^{-1} -2) 
    +16 \sqrt{-\Upsilon} \right] 
  F_{\Upsilon I} = 0 \;.
\end{eqnarray}
Upon variation of the entropy function with respect to $\Upsilon$ the
resulting equation is only covariant with respect to electric/magnetic
duality provided the attractor equations \eqref{eq:variational-y} are
satisfied. However, one can apply a mixed derivative of the form
$\delta = \partial/\partial_{\Upsilon} +\mathrm{i} F_{\Upsilon I} \,
N^{IJ} \, \partial/\partial Y^I$, which has the property that when
acting on a symplectic function $G(Y, \Upsilon)$, then also $\delta G$
transforms as a symplectic function \cite{deWit:1996ix}. An
alternative derivation is based on $\delta= Y^I\partial/\partial Y^I+
\bar Y^I\partial/\partial \bar Y^I+ 2 \Upsilon\partial/\partial
\Upsilon$, using that $\Upsilon$ is real so that
$\partial/\partial\Upsilon$ acts on both $\Upsilon$ and $\bar
\Upsilon$. Exploiting the homogeneity properties of the various
quantities involved, one derives the equation,
\begin{eqnarray}
  \label{eq:homo-delta}
  &&
  U\Sigma - \mathrm{i} (\bar Y^I F_I -  Y^I \bar F_I) \left[U + 4 
  (-\Upsilon)^{-1/2} (U-1) \right]\nonumber\\
  &&
  +2\mathrm{i}U  \left[ \Upsilon F_{I\Upsilon} N^{IJ} (\mathcal{Q}_J -
  \bar F_{JK}\mathcal{P}^K) - \mathrm{h.c.} \right] \nonumber \\
  &&
  +2\mathrm{i} (F_\Upsilon - \bar F_\Upsilon) \left[ 2 U\Upsilon + 4
  \sqrt{-\Upsilon} (1+U)\right]  = 0\;. 
\end{eqnarray}
Note that the above equations \eqref{eq:variational-y} and
\eqref{eq:homo-delta} are indeed satisfied in the BPS case. They are
also consistent with electric/magnetic duality.

\section{Discussion}
\label{sec:discussion}
\setcounter{equation}{0}
In this paper we studied the entropy function for static extremal
black holes using the proposal of \cite{Sen:2005wa} and we exhibited
its relation with the entropy function for BPS black holes in $N=2$
supergravity, derived in \cite{LopesCardoso:2006bg}. For BPS black
holes these two entropy functions lead to the same results for the
attractor equations and the entropy. This result even persists in the 
semi-classical approximation when evaluating an inverse Laplace
integral of the OSV-type \cite{LopesCardoso:2006bg}. 

In this final section we would like to discuss two more issues. The
first one deals with the presence of higher-derivative couplings other
than those introduced in section \ref{sec:N=2sugra}. The latter are
associated with interactions quadratic in the Riemann tensor and are
encoded by the $\Upsilon$ dependence in the holomorphic function
$F(Y,\Upsilon)$. Take, for instance, the simple example based on 
\begin{equation}
  \label{eq:example1}
  F(Y, \Upsilon) = - \frac{Y^1 Y^2 Y^3}{Y^0} - C \, \frac{Y^1}{Y^0} \,
  \Upsilon \;.  
\end{equation}
For BPS black holes the attractor equations can be solved for generic
charges \cite{LopesCardoso:1999ur}, but solutions only are only
consistent when the charges satisfy certain relations in which case
one obtains an explicit expression for the entropy. These relations
are not satisfied when the black hole carries the following
non-vanishing charges,
\begin{equation}
  \label{eq:example-charges}
  q_0 = p^1 = Q \;,\qquad p^2 = p^3 = P \;,
\end{equation}
with $PQ$ positive.  However, in that case \cite{Sahoo:2006rp},
non-supersymmetric black holes are possible and one can attempt to
solve the equations \eqref{eq:valueU}, \eqref{eq:variational-y} and
\eqref{eq:homo-delta}.  Unfortunately explicit solutions do not exist
and one has to resort to perturbation theory in the constant $C$.
To first order in $C$, the attractor values read, 
\begin{equation} 
  \label{eq:ysnon}
  \begin{array}{rcl}
    Y^0 &=& \ft14 P \left( 1  + 96\,C\,P^{-2} \right)\,,\\[.5ex]
    Y^1 &=& \ft14 \mathrm{i} \, Q \left(1+ 40 \,C\,P^{-2}\right)
    \,,\\[.5ex] 
    Y^2 &=& Y^3 = \,\ft14 \mathrm{i} \, P\left( 1
    +16\,C\,P^{-2}\right) \,, 
   \end{array} 
   \qquad 
  \begin{array}{rcl}
    ~&~& \\[.5ex] 
    U &=& 1 -  16\,C\,P^{-2}\,,\\[.5ex]
    \Upsilon &=& -4 \,.
   \end{array} 
\end{equation} 
In this order of perturbation theory the corresponding entropy
\eqref{eq:entropy<function} is computed by substituting the tree-level
values for $U$, $\Upsilon$ and the $Y^I$ into the entropy function
\eqref{eq:entropy-total-2}. The result reads,
\begin{equation}
  \label{eq:entrohyp}
  {\cal S}_{\rm macro} = 2\pi PQ  \left(1 + 40\, C \,P^{-2}\right)
  \;. 
\end{equation} 
As was argued in \cite{Sahoo:2006rp} this is not the expected value
from microstate counting \cite{Kraus:2005vz,Kraus:2005zm}, which
requires a different numerical factor in front of the $CP^{-2}$
correction term.  However, one has to take into account that other
higher-derivative interactions may be present, associated with matter
multiplets instead, which would in principle contribute to the
entropy. Such higher-derivative interactions have been studied for
$N=2$ tensor supermultiplets, and, indeed, it turns out that they lead
to entropy corrections for non-supersymmetric black holes
\cite{deWit:2006gn}. For BPS black holes, however, these corrections
vanish. Although a comprehensive treatment of higher-derivative
interactions is yet to be given for $N=2$ supergravity, it seems that
this result is generic.

These observations are in line with more recent findings
\cite{Exirifard:2006qv,Sahoo:2006pm} based on heterotic string
$\alpha^\prime$-corrections encoded in a higher-derivative effective
action in higher dimensions, which lead to additional matter-coupled
higher-derivative interactions in four dimensions.  When these are taken
into account, the matching of the macroscopic entropy with the
microscopic result is established \cite{Sahoo:2006pm}.

A second topic concerns possible non-holomorphic corrections to the
results presented in section \ref{sec:N=2sugra}. The Lagrangian
\eqref{eq:efflag} is based on a holomorphic homogeneous function
$F(X,\hat A)$, which subsequently is written in terms of the variables
$Y^I$ and $\Upsilon$, and corresponds to the so-called effective
Wilsonian action. This action is based on integrating out the massive
degrees of freedom and it describes the correct physics for energy
scales between appropriately chosen infrared and ultraviolet cutoffs.
In order to preserve physical symmetries non-holomorphic contributions
should be included associated with integrating out massless degrees of
freedom. In the special case of heterotic black holes in $N=4$
supersymmetric compactifications, the requirement of explicit
S-duality invariance of the entropy and the attractor equations allows
one to determine the contribution from these non-holomorphic
corrections, as was first demonstrated in \cite{LopesCardoso:1999ur}
for BPS black holes. In \cite{Cardoso:2004xf,LopesCardoso:2006bg} it
was established that non-holomorphic corrections to the BPS entropy
function \eqref{eq:Sigma-simple} can be encoded into a real function
$\Omega(Y,\bar Y,\Upsilon,\bar\Upsilon)$ which is homogeneous of
second degree. The modifications to the entropy function are then
effected by substituting $F(Y,\Upsilon)\to F(Y,\Upsilon) + 2\mathrm{i}
\Omega(Y,\bar Y,\Upsilon,\bar\Upsilon)$. There are good reasons to
expect that this same substitution should be applied to the more
general entropy function \eqref{eq:entropy-total-2}. Indeed, when
applying this ansatz to heterotic black holes in $N=4$ supersymmetric
compactifications, the resulting entropy function is S-duality
invariant and can be used to analyze non-supersymmetric extremal black
holes in the same way as was done for the BPS black holes. In that
case $\partial_\Upsilon(F +2\mathrm{i}\Omega)$ has to be an S-duality
invariant function. 

Unlike the BPS entropy function \eqref{eq:Sigma-simple}, the entropy
function \eqref{eq:entropy-total-2} was derived directly from an
effective action. Hence one may reconsider the relevant parts of this
effective action given in \eqref{eq:efflag-1-2}, in order to see
whether additional changes beyond the substitution 
 $F(Y,\Upsilon)\to F(Y,\Upsilon) + 2\mathrm{i}
\Omega(Y,\bar Y,\Upsilon,\bar\Upsilon)$
are needed in order to reproduce the
conjectured non-holomorphic modification of the entropy function. As
it turns out only one minor change is required. Namely, one has to
replace the coefficient $(F -F_IX^I +\ft12 \bar F_{IJ} X^IX^J)$ of the
$(T_{abij}\varepsilon^{ij})^2$ term in $\mathcal{L}_2$ by $(\hat A F_A
-\ft12 F_IX^I +\ft12 \bar F_{IJ} X^IX^J)$. For a holomorphic function
$F(X,{\hat A})$ these two expressions coincide by virtue of
\eqref{eq:homo-F}, but when the non-holomorphic function
$\Omega(X,{\bar X},{\hat A},\bar{\hat A})$ is included, the two
expressions will be different. Of course, the presence of
non-holomorphic terms will affect the supersymmetry of the original
action. Since the non-holomorphic corrections are expected to capture
the contributions of the massless modes, one expects that their
supersymmetrization will contain non-local interactions. The
construction of such a supersymmetric action is a challenge.

\subsection*{Acknowledgements}
We thank J. Perz for valuable discussions.  S.M. would like to thank
H. Nicolai and the members of quantum gravity group at AEI, Potsdam,
where part of this work was done, for the warm hospitality and
excellent academic atmosphere. This work is partly supported by EU
contracts MRTN-CT-2004-005104 and MRTN-CT-2004-512194, and by INTAS
contract 03-51-6346.

\begin{thebibliography}{10}
\bibitem{Ferrara:1995ih}
S.~Ferrara, R.~Kallosh and A.~Strominger, ``$N = 2$ extremal black holes'',
  {\em Phys. Rev.} {\bf D52} (1995) 5412--5416,
\href{http://www.arXiv.org/abs/hep-th/9508072}{{\tt hep-th/9508072}}.
%
\bibitem{Strominger:1996kf}
A.~Strominger, ``Macroscopic {entropy} of $N = 2$ {extremal} {black} {h}oles'',
  {\em Phys. Lett.} {\bf B383} (1996) 39--43,
\href{http://www.arXiv.org/abs/hep-th/9602111}{{\tt hep-th/9602111}}.
%
\bibitem{Ferrara:1996dd}
S.~Ferrara and R.~Kallosh, ``Supersymmetry and {a}ttractors'', {\em Phys. Rev.}
  {\bf D54} (1996) 1514--1524,
\href{http://www.arXiv.org/abs/hep-th/9602136}{{\tt hep-th/9602136}}.
%
\bibitem{Ferrara:1996um}
S.~Ferrara and R.~Kallosh, ``{Universality of supersymmetric attractors}'',
  {\em Phys. Rev.} {\bf D54} (1996) 1525--1534,
  \href{http://arXiv.org/abs/hep-th/9603090}{{\tt hep-th/9603090}}.
%
\bibitem{Ferrara:1997tw}
S. Ferrara, G.W. Gibbons and R. Kallosh,
``Black holes and critical points in moduli space'',
{\em Nucl. Phys.} {\bf B500} (1997) 75-93,
\href{http://www.arXiv.org/abs/hep-th/9702103}{{\tt hep-th/9702103}}.
%
\bibitem{Gibbons:1997cc}
G.W. Gibbons, 
``Supergravity vacua and solitons", in:
{\it Duality and Supersymmetric Theories}, eds. D.I. Olive and P.C. West,
Cambridge (1997) 267-296.
%
\bibitem{Behrndt:1996jn} K. Behrndt, G.L. Cardoso, B. de Wit,
R. Kallosh, D. L\"ust and T. Mohaupt,
``Classical and quantum {$N=2$} supersymmetric black holes'',
{\em Nucl. Phys.} {\bf B488} (1997) 236-260,
 {\tt hep-th/9610105}.
%
\bibitem{Denef:2000nb}
F.~Denef, ``{Supergravity flows and D-brane stability}'',  {\em JHEP} {\bf 08}
  (2000) 050, \href{http://arXiv.org/abs/hep-th/0005049}{{\tt
  hep-th/0005049}}.
%
\bibitem{LopesCardoso:1998wt}
G.L. Cardoso, B.~de~Wit and T.~Mohaupt, ``Corrections to macroscopic
  supersymmetric black-hole entropy'', {\em Phys. Lett.} {\bf B451} (1999)
  309--316,
\href{http://www.arXiv.org/abs/hep-th/9812082}{{\tt hep-th/9812082}}.
%
\bibitem{LopesCardoso:1999ur}
G.L. Cardoso, B.~de~Wit and T.~Mohaupt, ``Macroscopic entropy formulae and
  non-holomorphic corrections for supersymmetric black holes'', {\em Nucl.
  Phys.} {\bf B567} (2000) 87--110,
\href{http://www.arXiv.org/abs/hep-th/9906094}{{\tt hep-th/9906094}}.
%
\bibitem{LopesCardoso:2000qm}
G.L. Cardoso, B.~de~Wit, J.~{K\"appeli} and T.~Mohaupt, ``Stationary {BPS}
  solutions in $N = 2$ supergravity with {$R^2$}-interactions'', {\em JHEP}
  {\bf 12} (2000) 019,
\href{http://www.arXiv.org/abs/hep-th/0009234}{{\tt hep-th/0009234}}.
%
\bibitem{LopesCardoso:2006bg}
G.L. Cardoso, B. de Wit, J. K\"appeli and T. Mohaupt,
``Black hole partition functions and duality",
{\it JHEP} {\bf 03} (2006) 074, {\tt hep-th/0601108}.
%
\bibitem{Sen:2005wa}
A. Sen, ``Black hole entropy function and the attractor mechanism in
higher derivative gravity'', {\em JHEP} {\bf 0509} (2005) 038,
\href{http://www.arXiv.org/abs/hep-th/0506177}{{\tt hep-th/0506177}}.
%
\bibitem{Goldstein:2005hq}
K. Goldstein, N. Iizuka, R.P. Jena and S.P. Trivedi,
``Non-supersymmetric attractors'',
{\it Phys. Rev.} {\bf D72} (2005) 124021,
\href{http://www.arXiv.org/abs/hep-th/0507096}{{\tt hep-th/0507096}}.
%
\bibitem{Sen:2005iz}
A. Sen, ``Entropy function for heterotic black holes'',
{\it JHEP} {\bf 0603} (2006) 008,
\href{http://www.arXiv.org/abs/hep-th/0508042}{{\tt hep-th/0508042}}.
%
\bibitem{Kallosh:2005ax}
  R. Kallosh, ``New attractors'',
{\it JHEP} {\bf 0512} (2005) 022,
\href{http://www.arXiv.org/abs/hep-th/0510024}{{\tt hep-th/0510024}}.
%
\bibitem{Tripathy:2005qp}
 P.K. Tripathy and S.P. Trivedi, ``Non-supersymmetric attractors in
string theory'', {\it JHEP} {\bf 0603} (2006) 022,
\href{http://www.arXiv.org/abs/hep-th/0511117}{{\tt hep-th/0511117}}.
%
\bibitem{Giryavets:2005nf}
A. Giryavets, ``New attractors and area codes'', {\it JHEP} {\bf 0603} 
(2006) 020,
\href{http://www.arXiv.org/abs/hep-th/0511215}{{\tt hep-th/0511215}}.
%
\bibitem{Prester:2005qs}
P. Prester,
``Lovelock type gravity and small black holes in heterotic string theory'',
{\it JHEP} {\bf 0602} (2006) 039,
{\tt hep-th/0511306}.
%
\bibitem{Alishahiha} M. Alishahiha and H. Ebrahim,
``Non-Supersymmetric Attractors and Entropy Function'', 
{\it JHEP} {\bf 0603} (2006) 003,
{\tt hep-th/0601016}.
%
\bibitem{Kallosh:2006bt}
R.~Kallosh, N.~Sivanandam and M.~Soroush, ``{The non-BPS black hole attractor
  equation}'',  {\em JHEP} {\bf 03} (2006) 060,
  \href{http://arXiv.org/abs/hep-th/0602005}{{\tt hep-th/0602005}}.
%
\bibitem{Chandrasekhar:2006kx} B.~Chandrasekhar, S.~Parvizi,
  A.~Tavanfar and H.~Yavartanoo, ``{Non-supersymmetric attractors in
    $R^2$ gravities}'', {\em JHEP} {\bf 08} (2006) 004,
  \href{http://arXiv.org/abs/hep-th/0602022}{{\tt hep-th/0602022}}.
%
\bibitem{Bellucci:2006ew} S.~Bellucci, S.~Ferrara and A.~Marrani,
  ``{On some properties of the attractor equations}'', {\em Phys.
    Lett.} {\bf B635} (2006) 172--179,
  \href{http://arXiv.org/abs/hep-th/0602161}{{\tt hep-th/0602161}}.
%
\bibitem{Kallosh:2006bx}
R.~Kallosh, ``{From BPS to non-BPS black holes canonically}'',
  \href{http://arXiv.org/abs/hep-th/0603003}{{\tt hep-th/0603003}}.
%
\bibitem{Sahoo:2006rp}
B. Sahoo and A. Sen,  ``Higher derivative corrections to
non-supersymmetric extremal black holes in N=2 supergravity'', 
\href{http://www.arXiv.org/abs/hep-th/0603149}{{\tt hep-th/0603149}}.
%
\bibitem{Alishahiha:2006jd}
M.~Alishahiha and H.~Ebrahim, ``{New attractor, entropy function and black
  hole partition function}'',  {\em JHEP} {\bf 11} (2006) 017,
  \href{http://arXiv.org/abs/hep-th/0605279}{{\tt hep-th/0605279}}.
%
\bibitem{Exirifard:2006qv}
G. Exirifard,
``The world-sheet corrections to dyons in the heterotic theory",
{\it JHEP} {\bf 10} (2006) 070,
{\tt hep-th/0607094}.
%
\bibitem{Sahoo:2006pm} B. Sahoo and A. Sen,
  ``$\alpha^\prime$-Corrections to extremal dyonic black holes in
  heterotic string theory", {\tt hep-th/0608182}.
%
\bibitem{Astefanesei:2006sy}
D.~Astefanesei, K.~Goldstein and S.~Mahapatra, ``{Moduli and (un)attractor
  black hole thermodynamics}'',  
\href{http://arXiv.org/abs/hep-th/0611140}{{\tt
  hep-th/0611140}}.
%
\bibitem{Dabholkar:2006tb}
A.~Dabholkar, A.~Sen and S.~Trivedi, ``{Black hole microstates and attractor
  without supersymmetry}'',  \href{http://arXiv.org/abs/hep-th/0611143}{{\tt
  hep-th/0611143}}.
%
%
\bibitem{Sen:2004dp}
A.~Sen, ``How does a fundamental string stretch its horizon?'',
{\em JHEP} {\bf 0505} (2005) 059,
\href{http://www.arXiv.org/abs/hep-th/0411255}{{\tt hep-th/0411255}}.
%
\bibitem{Hubeny:2004ji}
V.~Hubeny, A.~Maloney and M.~Rangamani, ``String-corrected black holes'',
{\em JHEP} {\bf 0505} (2005) 035,
\href{http://www.arXiv.org/abs/hep-th/0411272}{{\tt hep-th/0411272}}.
%
\bibitem{Bak:2005mt}
D.~Bak, S.~Kim and S.-J. Rey, ``{Exactly soluble BPS black holes in higher
  curvature $N = 2$ supergravity}'',
  \href{http://arXiv.org/abs/hep-th/0501014}{{\tt hep-th/0501014}}.
%
\bibitem{Wald:1993nt} R.M. Wald, ``Black hole entropy entropy is
  Noether charge'', {\it Phys. Rev.} {\bf D48} (1993) 3427, {\tt
    gr-qc/9307038}; V. Iyer and R.M. Wald, ``Some properties of
  Noether charge and a proposal for dynamical black hole entropy'',
  {\it Phys. Rev.}  {\bf D50} (1994) 846, {\tt gr-qc/9403028};
  T. Jacobson, G. Kang and R.C. Myers, ``Black hole entropy in higher
  curvature gravity'', {\it Phys. Rev.} {\bf D49} (1994) 6587, {\tt
    gr-qc/9312023}.
%
\bibitem{Gaillard:1981rj}
M.K. Gaillard and B.~Zumino, ``{Duality rotations for interacting fields}'',  
{\it Nucl. Phys.} {\bf B193} (1981) 221.
%
\bibitem{deWit:2001pz}
B.~de~Wit, ``{Electric-magnetic duality in supergravity}'',  {\em
  Nucl. Phys. Proc.  Suppl.} {\bf 101} (2001) 154--171,
{{\tt hep-th/0103086}}.
%
\bibitem{deWit:1996ix} B. de Wit, ``{$N=2$} Electric-magnetic
duality in a chiral background'', 
{\em Nucl. Phys. Proc. Suppl.} {\bf 49}
(1996) 191-200, {\tt hep-th/9602060};
``{$N=2$} symplectic reparametrizations in a chiral background'',
{\em Fortschr. Phys.} {\bf 44} (1996) 529-538,
 {\tt hep-th/9603191}.
%
\bibitem{deWit:1982na}
B.~de Wit, R.~Philippe and A.~Van Proeyen,
    ``The improved tensor multiplet in N=2 supergravity'',
    {\em Nucl.\ Phys.}\  {\bf B219} (1983) 143.
%
\bibitem{dWKV:1999}
    B.~de Wit, B.~Kleijn and S.~Vandoren,
    ``Superconformal hypermultiplets'',
    {\it Nucl.\ Phys.}\ {\bf B568} (2000) 475, {\tt hep-th/9909228}.
%
\bibitem{Cardoso:2006cb} G.L. Cardoso, V.~Grass, D.~L\"ust and
  J.~Perz, ``{Extremal non-BPS black holes and entropy
    extremization}'', {\em JHEP} {\bf 09} (2006) 078,
  \href{http://arXiv.org/abs/hep-th/0607202}{{\tt hep-th/0607202}}.
%
\bibitem{Ooguri:2004zv}
H.~Ooguri, A.~Strominger and C.~Vafa, ``Black hole attractors and the
  topological string'', {\em Phys. Rev.} {\bf D70} (2004) 106007,
\href{http://www.arXiv.org/abs/hep-th/0405146}{{\tt hep-th/0405146}}.
%
\bibitem{Kraus:2005vz}
P. Kraus and F. Larsen,
``Microscopic black hole entropy in theories with higher derivatives",
{\it JHEP} {\bf 09} (2005) 034, 
{\tt hep-th/0506176}.
%
\bibitem{Kraus:2005zm} P. Kraus and F. Larsen, ``Holographic
  gravitational anomalies", {\it JHEP} {\bf 01} (2006) 022, {\tt
    hep-th/0508218}.
%
\bibitem{deWit:2006gn}
  B.~de Wit and F.~Saueressig,
  ``{Off-shell N = 2 tensor supermultiplets}'',
{\it JHEP} {\bf 0609} (2006) 062, 
  {\tt hep-th/0606148}.
%
\bibitem{Cardoso:2004xf} G.L. Cardoso, B. de Wit, J. K\"appeli and T. Mohaupt,
  ``Asymptotic degeneracy of dyonic $N=4$ string states and black hole
  entropy'', {\it JHEP} {\bf 12} (2004) 075, {\tt hep-th/0412287}.
%
\end{thebibliography}
\providecommand{\href}[2]{#2}
\begingroup\raggedright\endgroup
\end{document}